\documentclass[aps,prl,amsmath,amssymb,twocolumn,floatfix,superscriptaddress,nofootinbib]{revtex4-1}
\usepackage{graphicx}
\usepackage{bm}
\usepackage{latexsym}
\usepackage{epsf}
\usepackage{rotating}
\usepackage{epsfig,graphics,rotate,color}
\usepackage{wrapfig}
\usepackage{gensymb}
\usepackage{amssymb}
\usepackage{amsmath}
\usepackage{amsfonts}
\usepackage{array,hhline,dcolumn}
\usepackage[normalem]{ulem}
\usepackage{color}
\usepackage{xcolor}
\usepackage{hyperref}
\usepackage{mwe}

\newcommand{\U}{$^{238}$U } 
\newcommand{\Be}{$^{10}$Be }

\begin{document}
\title{Measuring Changes in the Atmospheric Neutrino Rate over Gigayear Timescales}
\author{Johnathon~R.~Jordan}
\email{jrlowery@umich.edu}
\affiliation{University of Michigan, 450 Church St, Ann Arbor, MI 48109, USA}
\author{Sebastian~Baum}
\email{sbaum@stanford.edu}
\affiliation{Stanford Institute for Theoretical Physics, Department of Physics, Stanford University, Stanford, CA 94305, USA}
\affiliation{The Oskar Klein Centre for Cosmoparticle Physics, \\Department of Physics, Stockholm University, Alba Nova, 10691 Stockholm, Sweden}
\author{Patrick~Stengel}
\email{patrick.stengel@fysik.su.se}
\affiliation{The Oskar Klein Centre for Cosmoparticle Physics, \\Department of Physics, Stockholm University, Alba Nova, 10691 Stockholm, Sweden}
\author{Alfredo~Ferrari}
\affiliation{CERN, 1211 Geneva 23, Switzerland}
\author{Maria~Cristina~Morone}
\affiliation{Physics Department, University of Roma Tor Vergata, 00133 Rome, Italy}
\affiliation{INFN Roma Tor Vergata, 00133 Rome, Italy}
\author{Paola~Sala}
\affiliation{INFN Milano, via Celoria 16, 20133 Milano, Italy}
\author{Joshua~Spitz}
\email{spitzj@umich.edu}
\affiliation{University of Michigan, 450 Church St, Ann Arbor, MI 48109, USA}

\date{\today}

%%%%%%%%%%%%%%%%%%%%%%%%%%%%%%%%%%%%%%%%%%%%%%%%%%%%%%%%%%%%
\begin{abstract}
Measuring the cosmic ray flux over timescales comparable to the age of the Solar System, $\sim 4.5\,$Gyr, could provide a new window on the history of the Earth, the Solar System, and even our Galaxy. We present a technique to indirectly measure the rate of cosmic rays as a function of time using the imprints of atmospheric neutrinos in {\it paleo-detectors}, natural minerals which record damage tracks from nuclear recoils. Minerals commonly found on Earth are $\lesssim 1\,$Gyr old, providing the ability to look back across cosmic ray history on timescales of the same order as the age of the Solar System. Given a collection of differently aged samples dated with reasonable accuracy, this technique is particularly well-suited to measuring historical {\it changes} in the cosmic ray flux at Earth and is broadly applicable in astrophysics and geophysics.
\end{abstract}
%%%%%%%%%%%%%%%%%%%%%%%%%%%%%%%%%%%%%%%%%%%%%%%%%%%%%%%%%%%%
\maketitle

%%%%%%%%%%%%%%%%%%%%%%%%%%%%%%%%%%%%%%%%%%%%%%%%%%%%%%%%%%%%
The sources, composition, and propagation of cosmic rays through our Galaxy remain some of the biggest mysteries in astrophysics despite a century of experimental effort using a variety of techniques. Today, balloon- and space-based experiments measure the flux of cosmic rays \textit{directly}, while terrestrial experiments probe the nature of cosmic rays \textit{indirectly} via measurements of secondary particles created by their interactions with the Earth's atmosphere. Notably, {\it atmospheric neutrinos} arising from such cascades can be detected in terrestrial experiments~\cite{Fukuda:2002uc,Aartsen:2016nxy,Aharmim:2009zm,Adamson:2012gt,AdrianMartinez:2012ph}.

Characterizing atmospheric neutrinos provides a window on a variety of physical processes, many of which cannot easily be studied by other means. The flux of primary cosmic rays incident on the atmosphere depends on the sources and composition of the cosmic rays themselves as well as their subsequent propagation through the cosmos. Upon reaching our Solar System, the cosmic ray trajectories are further altered by the magnetic fields of the Sun and Earth~\cite{Lipari:2000wu,Battistoni:2005,Driscoll:2016,Muscheler:2016}. Finally, the composition and structure of the Earth's atmosphere impact the observed neutrino flux at the surface~\cite{Battistoni:2002ew,Battistoni:2005,Chemke:2017}.

{\it Paleo-detectors} are natural minerals which can record and retain tracks formed by nuclear recoils induced by atmospheric neutrino interactions. We propose using a series of paleo-detectors, with ages $\sim$10$^8 - 10^9$\,years dated to few-percent accuracy~\cite{GTS2012,Gallagher:1998,vandenHaute:1998}, to measure changes in the atmospheric neutrino rate over geological timescales. Previously, paleo-detectors have been proposed to measure interactions of neutrinos from galactic supernovae~\cite{Baum:2019fqm} and dark matter~\cite{Baum:2018tfw,Drukier:2018pdy,Edwards:2019puy}. Recent advances in a variety of read-out techniques potentially allow for macroscopic samples ($\sim 100\,$g) of target material to be imaged with nano-scale resolution in three dimensions, see the discussion in~\cite{Baum:2018tfw,Drukier:2018pdy} and~\cite{RODRIGUEZ2014150,BARTZ2013273,Kouwenberg:2018,NORDLUND2018450}, enabling much larger exposure and better characterization of the track length distributions than previous ideas to probe rare events using natural minerals~\cite{Goto:1958,Goto:1963zz,Fleischer:1969mj,Fleischer:1970zy,Fleischer:1970vm,Alvarez:1970zu,Kolm:1971xb,Eberhard:1971re,Ross:1973it,Price:1983ax,Kovalik:1986zz,Price:1986ky,Ghosh:1990ki,Jeon:1995rf,Collar:1999md,SnowdenIfft:1995ke,Collar:1994mj,Engel:1995gw,SnowdenIfft:1997hd}.

Compared to other tracers of the primary cosmic ray flux, e.g. cosmogenic muons, atmospheric neutrinos have the advantage that they are not screened by the Earth. The cosmogenic muon flux is exponentially sensitive to the height of the overburden; thus, using muon-induced recoils to infer the cosmic ray flux with paleo-detectors would require exquisite knowledge of the sample's geological history. Using atmospheric-neutrino-induced recoils, the required geological knowledge is only that samples have been buried deeper than $\sim 5\,$km, such that cosmogenic muons are sufficiently shielded (see~\cite{Baum:2019fqm,Baum:2018tfw,Drukier:2018pdy} for discussion). Mineral samples from such depths can be obtained from bore-hole cores typically used for geology research and oil exploration (see, e.g.,~\cite{Blattlereaar2687,Hirschmann1997,KREMENETSKY198611}). Paleo-detector measurements of the atmospheric neutrino flux in samples from deep bore-holes can thus be sensitive to the evolution of the cosmic ray flux over geological timescales.

To date, this evolution has only been measured by studying rare isotopes produced in cosmic ray interactions. For example, the deposition of cosmogenic nuclides, such as \Be, in the Earth's crust allow us to track the evolution of the cosmic ray flux and Earth's atmosphere going back as far as $\sim 10\,$Myr~\cite{Lal:1967,Wieler:2013}. Cosmogenic nuclides in meteorites can include radiogenic isotopes with much longer half-lives, (e.g.\ $^{40}$K with $T_{1/2} \sim 1.2\,$Gyr), providing information about the galactic cosmic ray flux on gigayear timescales (for a review, see~\cite{Wieler:2013}). However, such studies infer the flux of cosmic rays at a given meteorite's location and are thus not applicable to studying local influences on the cosmic ray flux at Earth.

Here, we consider the capability of paleo-detectors to differentiate several generic scenarios of cosmic ray flux evolution without regard for the specific underlying physics - such changes could be gradual (e.g.\ originating from slowly evolving star formation or supernova rates) or transient (e.g.\ caused by a nearby supernova or merger of neutron stars). For definiteness, we will consider halite (NaCl) samples, but our conclusions hold for other target materials. To get a sense of the expected signal rate, we can estimate the total atmospheric neutrino flux as $\sim1$~cm$^{-2}$~s$^{-1}$ and the neutrino-nucleon interaction cross section to be $\sim 10^{-38}\,{\rm cm}^2$. In a 1\,Gyr old 100\,g sample, this yields a total of $\sim 10^4$ neutrino interactions.

As we will discuss below, the interactions of atmospheric neutrinos also produce secondary particles (e.g. neutrons) which, in turn, can induce additional nuclear recoils. Thus, the total number of recoils will be a few times higher than the neutrino interaction rate. Using a full simulation and taking into account relevant backgrounds, we will show that with the present-day neutrino flux, atmospheric neutrino interactions create $\sim 6 \times 10^4$ tracks from nuclear recoils in the background-free signal region of a 1\,Gyr old 100\,g sample. Given this large number of signal tracks and the ability to control systematics with multiple mineral samples, we anticipate that paleo-detectors could be highly sensitive to changes in the cosmic ray flux over $\sim 1\,$Gyr.

%%%%%%%%%%%%%%%%%%%%%%%%%%%%%%%%%%%%%%%%%%%%%%%%%%%%%%%%%%%%
{\it Atmospheric Neutrino Signal.}---Before assessing the sensitivity of paleo-detectors to time-varying effects, we discuss the modeling of the atmospheric neutrino signal. For simplicity, we assume that the shape of the atmospheric neutrino spectrum is constant in time. Changes in the spectral shape would affect the expected signal, but for the simple counting experiment we propose, these changes just alter the observed rate in the signal regions and hence would be largely indistinguishable from a change in overall normalization. A spectral analysis could exploit the track length distribution's dependence on the shape of the atmospheric neutrino flux; however, the nuclear recoil spectrum is only weakly dependent on the primary cosmic ray spectrum because of the chain of intermediate interactions.

We calculate the neutrino flux with the multi-particle transport and interaction code \texttt{FLUKA}~\cite{Bohlen:2014buj,Ferrari:2005zk,NUNDIS} as in~\cite{Battistoni:2002ew} for minimum solar modulation, at the location of the Gran Sasso underground laboratory (42.44$^\circ$N, 13.57$^\circ$E). 

The signal tracks in a paleo-detector arise from the interactions of the atmospheric neutrino with the target material. We model signal tracks in halite from recoils of the constituent nuclei as well as lighter nuclei produced by inelastic neutrino interactions with the target. Secondary nuclear recoils are induced by products of neutrino interactions (in particular neutrons, protons, pions, muons, and kaons) with the target. As we will see, the most relevant signal tracks have track lengths $x \gtrsim 1\,\mu$m. About 75\% of such tracks stem from the interactions of secondary particles. We note that the neutrinos producing such secondary particles have typical energies $\gtrsim 100\,$MeV, and the associated energy of the primary cosmic rays producing such neutrinos in the atmosphere is $\gtrsim 10\,$GeV. 

We model all neutrino interactions and particle propagation through the target material with \texttt{FLUKA}, which implements quasi-elastic, resonant ($\Delta$ production only), and deep inelastic scattering (DIS) interactions along with initial- and final-state effects. We consider all nuclear recoil tracks stemming from interactions of atmospheric neutrinos with the target nuclei to be signal. Accurate modeling of neutrons in the target material is crucial to predicting both the signal component from neutrons produced in neutrino interactions and the background from radiogenic neutrons (discussed below). Details on the treatment of neutrons in previous versions of \texttt{FLUKA} can be found in \cite{Ferrari:2005zk}. For this work, an unreleased version of \texttt{FLUKA} containing new algorithms for the interactions of low energy neutrons is used. The new treatment ensures energy conservation for each interaction and provides an estimate of the recoil energy. These algorithms are still under development and are thus only available for a subset of isotopes. In particular, the cross sections for Cl are not implemented and so a similar nucleus (P) is substituted.

\begin{figure}
    %\includegraphics[width=0.49\linewidth]{recoilEnergyVsA}
    %\hfill
    \includegraphics[width=\linewidth]{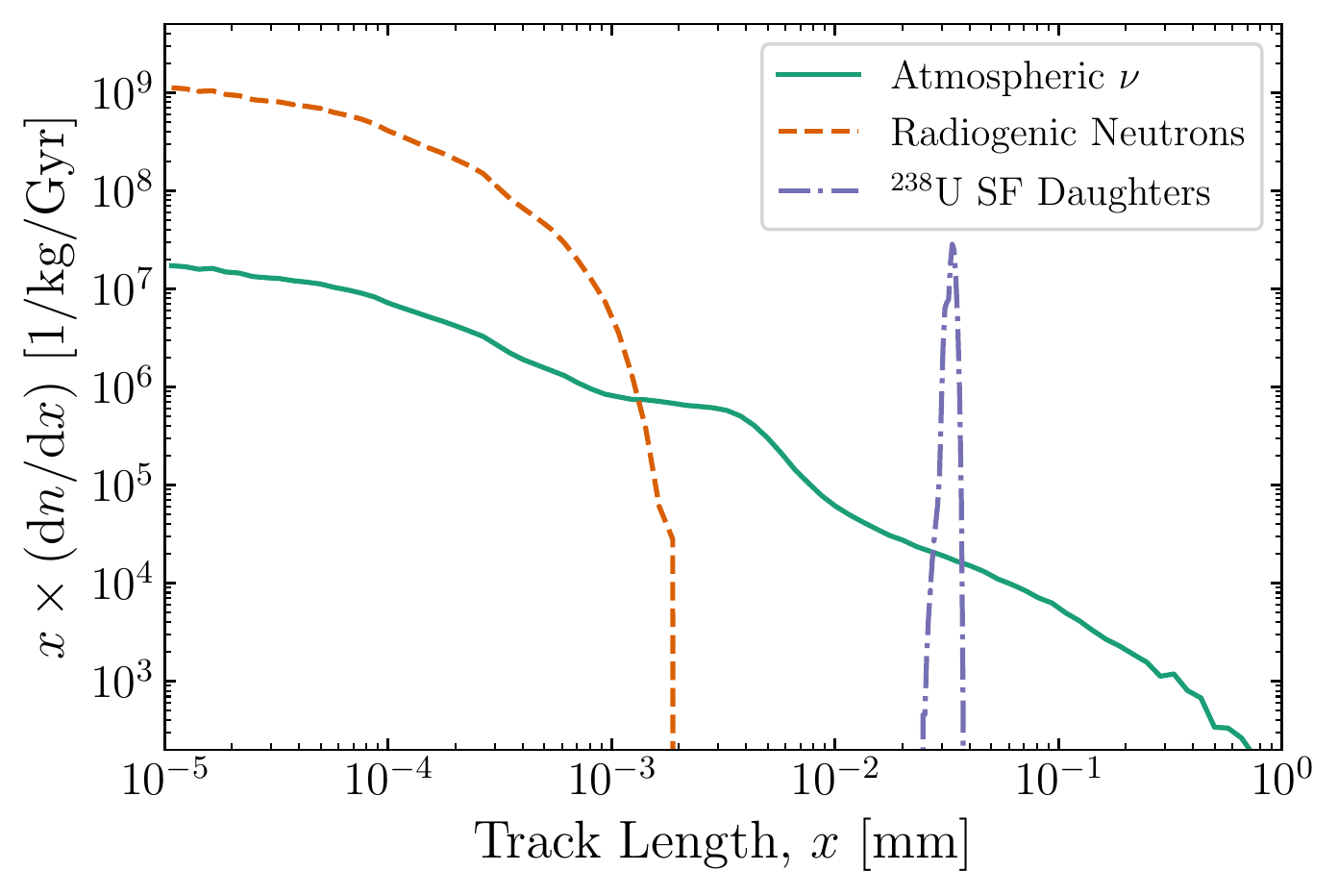}
    \caption{Differential distribution of the number of recoils per unit target mass and unit time ($n$) with respect to the recoil track length ($x$) for a halite paleo-detector (modeled as $^{23}$Na and $^{31}$P; see text). In addition to the signal induced by interactions of atmospheric neutrinos (solid green), we also show the spectrum of background tracks induced by interactions of radiogenic neutrons (dashed orange) and the track length spectrum from spontaneous fission daughters (purple dash-dotted) assuming a $^{238}$U concentration of 0.01 ppb by weight, consistent with previous work on paleo-detectors~\cite{Baum:2019fqm}. See the text for a discussion of these backgrounds. Note that there are two virtually background-free track length regions: $2\,\mu{\rm m} \leq x \leq 20\,\mu$m and $50\,\mu{\rm m} \leq x \leq 1\,$mm. Only recoils with $A>4$ are included; lighter nuclei are not expected to give rise to visible tracks.}
    \label{FIG:RecoilProperties}
\end{figure}

We simulate the neutrino interactions and radiogenic backgrounds at the center of a large (20\,m)$^3$ uniform cube to ensure that all products from atmospheric neutrino interactions are contained in the simulated volume. For each recoil in \texttt{FLUKA}, we convert from recoil energy to track length using stopping powers from \texttt{SRIM}~\cite{Ziegler:1985,Ziegler:2010}; we show the resulting track length spectra for both the atmospheric neutrino signal and the radiogenic backgrounds (discussed below) in Fig.~\ref{FIG:RecoilProperties}.

We assume the same read-out benchmark previously used in studies of paleo-detectors for dark matter and core collapse supernova neutrinos. By imaging with small angle X-ray scattering at a synchrotron facility, we estimate that a $\sim 100 \,$g sample could be read out with three-dimensional spatial resolution of $\sim 15 \,$nm~\cite{Holler2014, Schaff2015}. We note that read-out techniques with lower spatial resolution and higher throughput could potentially be used to image the higher energy recoil tracks investigated in this work (see~\cite{Drukier:2018pdy} for discussion).

The atmospheric neutrino flux depends on a variety of physical systems which all evolve independently over geological timescales. Cosmogenic nuclide studies of meteorites suggest that the primary cosmic ray intensity in our Galaxy has increased by a factor of $\sim 1.5$ over the last $\sim 1\,$Gyr~\cite{Wieler:2013}. While the propagation of cosmic rays through our Solar System depends on the strength and orientation of the heliomagnetic field, such effects are modulated much faster than the $\sim 100\,$Myr geological timescales relevant for paleo-detectors (see~\cite{Muscheler:2016} and references within).

Cosmic rays are also deflected by the geomagnetic field before interacting with the Earth's atmosphere. The geomagnetic field is well-approximated by a dipole and studies of the geodynamo and paleomagnetic records (for example, see~\cite{Driscoll:2016}) suggest that the dipole moment could have varied by an ${\cal O}(1)$ factor over Gyr timescales. The main effect of the geomagnetic field on the primary cosmic ray spectrum hitting the atmosphere is a {\it rigidity cutoff}, providing a lower-energy limit on primary cosmic rays, and in turn, the associated atmospheric neutrino spectra. The rigidity cutoff depends on the location of the target mineral relative to the orientation of the geomagnetic fields and is directly proportional to the dipole moment (for example, see~\cite{Lipari:2000wu}); we will discuss the implications for paleo-detectors below.

Finally, the atmospheric neutrino flux depends on the composition and density of the atmosphere. The change in atmospheric composition over the relevant timescales $\lesssim 1\,$Gyr considered here is primarily the replacement of a fraction of the ${\rm N}_2$ with ${\rm O}_2$ (for example, see~\cite{JOHNSON2015150}), which does not significantly alter the cascades that yield atmospheric neutrinos. A variety of studies (see~\cite{Chemke:2017} and references within) suggest that the density of the atmosphere could have varied by as much as a factor of two downwards or an order of magnitude upwards going back $\mathcal{O}(1)\,$Gyr. Since the atmosphere today is thicker than ten interaction lengths of a typical cosmic ray proton~\cite{Battistoni:2002ew,Battistoni:2005}, we expect no significant modification of the atmospheric neutrino flux from such density changes.

%%%%%%%%%%%%%%%%%%%%%%%%%%%%%%%%%%%%%%%%%%%%%%%%%%%%%%%%%%%%
{\it Radiogenic Backgrounds.}---A large number of background tracks from $\alpha$-particles produced in the decay chains of radioactive contaminants such as \U could potentially impact the sensitivity of paleo-detectors to atmospheric neutrinos. However, very light nuclei ($A < 5$, i.e. H and He) generally have stopping powers too small to leave robust tracks in minerals (e.g. see discussion in~\cite{Drukier:2018pdy} and references therein), although some recent work has shown that it is possible to read out $\alpha$-particle tracks in some materials~\cite{BARTZ2013273,Kouwenberg:2018}. Here, we assume that recoiling nuclei with $A < 5$ do not leave resolvable tracks. For materials and readout technologies where such tracks are visible, the qualitative results of this paper will remain the same.

As discussed in previous work~\cite{Baum:2019fqm,Baum:2018tfw,Drukier:2018pdy,Edwards:2019puy}, tracks from the heavy nuclear remnants of $\alpha$-decays are significantly shorter than the track lengths relevant for the sensitivity to atmospheric neutrinos. The most important remaining backgrounds from \U come in two forms: 1) neutrons produced in radioactive decays which scatter off target nuclei and 2) daughter nuclei from \U spontaneous fission (SF) which are generally much heavier and more energetic than the nuclei created by atmospheric neutrino interactions with halite. Note that the tracks from SF daughters can be used for fission track dating of the samples.

We calculate the primary neutron spectra from the entire $^{238}{\rm U} \to \ldots \to {^{206}{\rm Pb}}$ decay chain with \texttt{SOURCES}~\cite{sources4a:1999}. Our calculation includes the neutrons produced by \U SF (and the nuclei in its decay chain) as well as neutrons produced in $(\alpha,n)$ reactions of \U decay chain $\alpha$-particles with Na and Cl. These neutrons lose their energy predominantly via elastic interactions with the nuclei comprising the target material, giving rise to a large number of relatively soft nuclear recoils. We propagate the neutrons through the material and calculate the neutron-induced recoil spectrum with \texttt{FLUKA}; the associated tracks dominate the track length spectrum at lengths $\lesssim 1\,\mu$m.

Daughters from \U SF are modeled with \texttt{FREYA}~\cite{Verbeke:2015hka}, which produces correlated fission secondaries using a combination of data and analytical models. The daughter nuclei from each fission event come out approximately back-to-back, so we treat the two daughter tracks as a single longer track. We conservatively assume that the chosen readout technique is insensitive to the distribution of energy loss along each track although this would in principle be useful for distinguishing different recoiling nuclei and identifying the two fission-daughter tracks. The SF daughters produce tracks with characteristic length $\sim 25-40\,\mu$m in halite.

%%%%%%%%%%%%%%%%%%%%%%%%%%%%%%%%%%%%%%%%%%%%%%%%%%%%%%%%%%%%
{\it Sensitivity to Atmospheric Neutrinos.}---In previous studies of paleo-detectors~\cite{Baum:2018tfw,Drukier:2018pdy,Edwards:2019puy,Baum:2019fqm}, which investigated the sensitivity to signals associated with keV-scale recoils, the dominant background was due to radiogenic neutrons. In Fig.~\ref{FIG:RecoilProperties}, we see that atmospheric neutrino interactions create tracks much longer than those arising from radiogenic neutrons. Consequently, the level of radiopurity in the sample does not strongly affect the sensitivity of the proposed measurement and the qualitative conclusions presented below should hold for similar minerals which can form and retain tracks on geological timescales.

Mineral samples of different ages should be associated with the same type of host rocks in order to help control for any potential systematics arising from different characteristic geological histories. For instance, the contribution of non-neutrino cosmic rays close to the surface (e.g.\ muons and neutrons) could vary significantly. Changes to the depth of host rocks with time could raise or lower the cosmogenic muon (and the associated neutron) flux. As discussed in~\cite{Baum:2018tfw,Baum:2019fqm}, many halite-bearing evaporite deposits feature salt structures which %become accessible to bore-hole drilling after intruding
intrude into the overlying rock. During this intrusion (called {\it diapirism}) the halite typically re-crystallizes, effectively erasing any latent tracks induced by cosmic rays during the initial burial period. We therefore simply consider cosmogenic muons/neutrons as a background which can be avoided by retrieving samples with a variety of ages from evaporite deposits at sufficient depths $\gtrsim 5 \,$km in different locations (for example, see~\cite{Evans:2006,Warren:2006}).

Systematics associated with flux variations as a function of either latitude or time arising from (necessarily) excavating different-age samples at different locations can potentially be constrained or measured in the background-free region of the track length spectrum at shorter track lengths. The tracks in the $2\,\mu{\rm m} \leq x \leq 20\,\mu$m signal region primarily arise from nuclei with energies $\lesssim 10 \,$MeV and masses close to or matching those of the nuclei comprising the target mineral. Such recoils typically arise in (quasi)elastic scattering of neutrinos or secondary neutrons off the target nuclei. The rigidities of cosmic ray particles capable of producing atmospheric neutrinos at energies where (quasi)elastic scattering dominates the interactions with the target are similar to the range of geomagnetic rigidity cutoffs one might expect at different latitudes today.

The signal tracks in the background-free region at longer track lengths, $50\,\mu{\rm m} \leq x \leq 1\,$mm, arise exclusively from nuclei with masses much lighter than those comprising the target. Such lighter nuclei are produced by deep inelastic scattering (DIS) of neutrinos with the target nuclei. Cosmic rays giving rise to atmospheric neutrinos sufficiently energetic to interact with the target via DIS have energies well above the largest rigidity cutoff one expects at any latitude, even after accounting for a potential ${\cal O}(1)$ variation in the geomagnetic field strength over the last Gyr. This signal region is thus largely independent of the systematics associated with the geomagnetic field.

\begin{figure}
    \includegraphics[width=\linewidth]{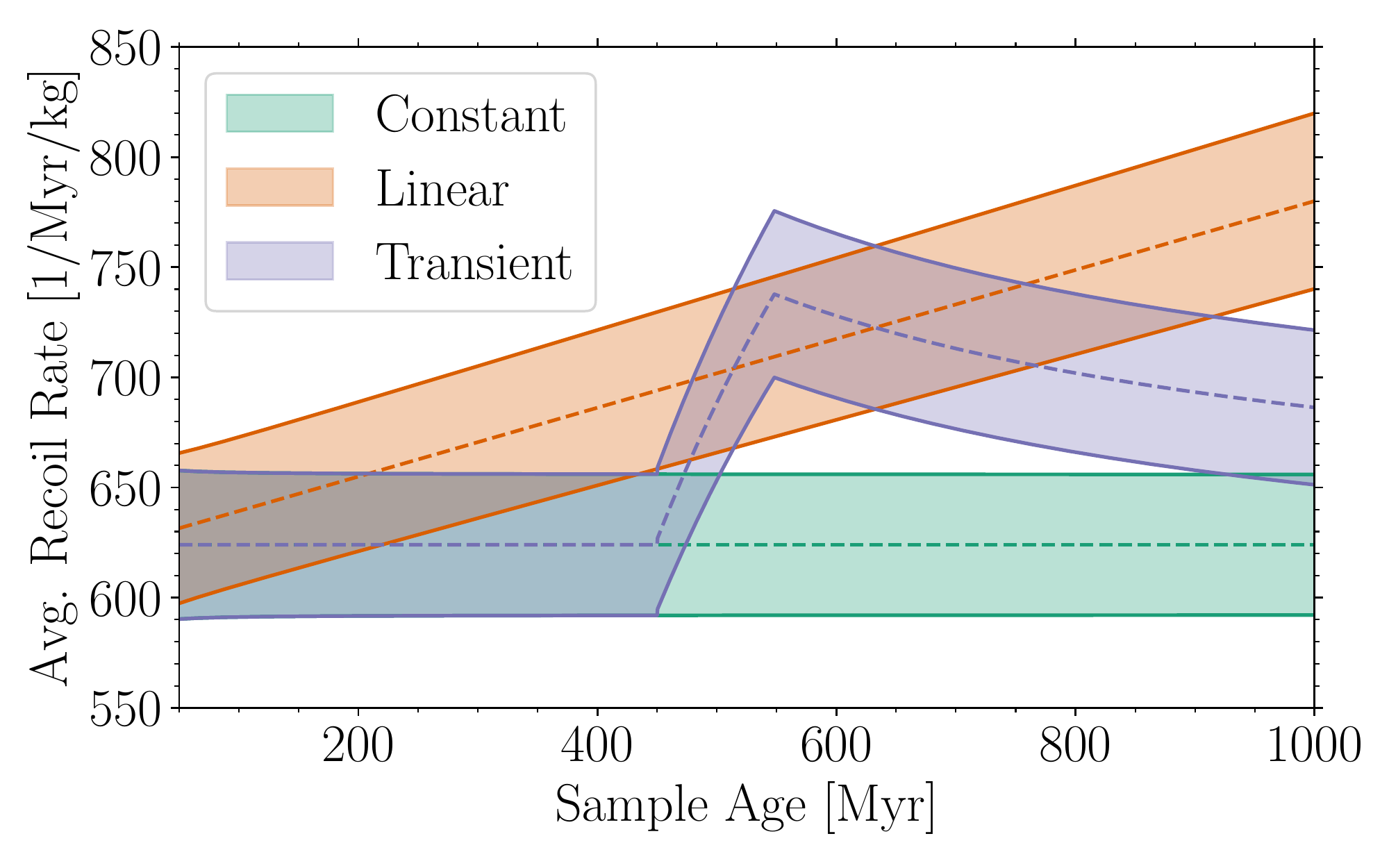}
    \caption{The average recoil rate one would infer from samples as a function of their age, for three different scenarios: constant atmospheric neutrino flux (green), linearly decreasing flux with a 50\% drop over 1\,Gyr (orange), and constant flux with a transient event where the flux is elevated by 100\% for 100 Myr (purple). Error bands represent the 1$\sigma$ error one would obtain on the average recoil rate from a sample of a given age, assuming that the age of samples can be determined with relative uncertainty of 5\% and its mass with 1\% uncertainty.}
    \label{FIG:cartoon}
\end{figure}

Here, we do not include effects of annealing or track fading. While detailed studies are required to determine a precise model for track formation and retention in the variety of possible paleo-detector minerals, tracks associated with the much higher energy nuclear recoils induced by atmospheric neutrinos should be more robust to annealing compared to tracks induced by dark matter or neutrinos from core collapse supernovae. Such studies have been carried out for recoils induced by SF daughters~\cite{Fleischer:1965yv,GUO2012233}, which produce tracks similar to the signal. Any effects of track fading measured in an analysis of the fission tracks could, in principle, provide for a template to model the fading of signal tracks. In any case, signal and background track fading should be similar and, thus, large signal-rich, background-free swaths of the track length spectrum will remain.

Measuring the time evolution of the atmospheric neutrino flux requires counting the number of atmospheric-neutrino-induced tracks $N_i$ in (the background free regions of) multiple mineral samples of different ages $t_i$. A simple quantity one could construct from such measurements is the average recoil rate per unit target mass over the age of the sample, $\overline{r}_i=N_i/t_i M_i$, where $M_i$ is the sample mass. Given the large number of tracks expected in a representative sample (see Fig.~\ref{FIG:RecoilProperties}), a good estimate for the relative error on $\overline{r}_i$ can be obtained by summing the relative systematic errors of the quantities entering the rate in quadrature. One source of systematic errors is the relation of the measured number of tracks $N_i$ and the true number of recoils induced by atmospheric neutrino interactions. Such errors could e.g. be induced via mis-modeling of backgrounds, track fading, or read-out inefficiencies. Further systematic errors are due to imperfect determination of the samples' ages and masses.

In Fig.~\ref{FIG:cartoon} we show a cartoon of the average recoil rate $\overline{r}$ as a function of the age of a sample for three different scenarios: constant atmospheric neutrino flux (equal to the present-day flux), linearly decreasing flux with a 50\% drop over 1\,Gyr, and a transient event where the flux is elevated by 100\% between 450\,Myr and 550\,Myr ago. To get an idea for the uncertainty on $\overline{r}$, we assume that the samples' ages and masses can be determined with relative precision of $\Delta_t = 5\%$ and $\Delta_M = 1\%$, and that such errors dominate over systematic errors on $N_i$. The shaded bands indicate the corresponding error estimate. Quantitative statements about differentiating various scenarios for the atmospheric neutrino flux history are dependent on the specific parameters of the excavated samples ($M_i, t_i$) and the statistical procedure used for a particular analysis, cf.~\cite{Baum:2019fqm}. However, Fig.~\ref{FIG:cartoon} indicates that, with the above error assumptions, a modest number of samples with different ages covering $t_i \sim 100\,{\rm Myr} - 1\,$Gyr would allow us to decisively differentiate scenarios which differ at least as much from each other as those shown in Fig.~\ref{FIG:cartoon}.

%%%%%%%%%%%%%%%%%%%%%%%%%%%%%%%%%%%%%%%%%%%%%%%%%%%%%%%%%%%%
{\it Conclusions.}---Paleo-detectors are a unique tool to study the evolution of the cosmic ray flux at Earth over gigayear timescales using the traces of nuclear recoils induced by atmospheric neutrino interactions. Due to the large exposures that can be achieved using natural minerals ($\sim0.1$\,kg\,Gyr) and the presence of background-free track length regions, paleo-detectors are well-suited to measuring changes in the atmospheric neutrino flux over time. In a simple analysis which counts the recoil tracks in two background-free track length regions (e.g. $2\,\mu{\rm m} \leq x \leq 20\,\mu$m and $50\,\mu{\rm m} \leq x \leq 1\,$mm in halite), the total number of recoils expected in a 100\,g, 1\,Gyr old sample is $\sim 6 \times10^4$.

We note that, while a simple counting experiment is unable to isolate the effects of the different physical mechanisms affecting the flux, progress towards this end is possible with more sophisticated analysis techniques. Combining measurements from a modest sample of ancient minerals of varying ages allows one to distinguish scenarios of cosmic ray flux evolution and constrain systematic uncertainties due to differences in sample composition, age, and location. These measurements offer a novel approach to studying the history of cosmic rays at Earth over \textit{gigayear} timescales. Paleo-detectors could thus open a new window into the evolution of the Earth, in addition to providing for a more complete picture of astrophysics in our Galaxy.

\begin{acknowledgments}
We thank J.~F.~Beacom, K.~Freese, and M.~W.~Winkler for helpful suggestions and comments.
S.~Baum and P.~Stengel thank the LCTP at the University of Michigan, where part of this work was completed, for hospitality. 
J.~R.~Jordan is supported by the National Science Foundation Graduate Research Fellowship under Grant No. DGE-1256260. 
J.~Spitz is supported by the Department of Energy, Office of Science, under Award No. DE-SC0007859. 
S.~Baum is supported in part by NSF Grant PHY-1720397, DOE HEP QuantISED award \#100495, and the Gordon and Betty Moore Foundation Grant GBMF7946. 
S.~Baum and P.~Stengel acknowledge support by the Vetenskapsr\r{a}det (Swedish Research Council) through contract No. 638-2013-8993 and the Oskar Klein Centre for Cosmoparticle Physics. 
P.~Sala thanks the FLUKA collaboration. 
\end{acknowledgments}

\bibliography{thebib}

\end{document}